\documentclass[amsmath,aps,twocolumn]{revtex4-2}
\usepackage{graphicx}
\usepackage{times}
\usepackage{slashed}
\usepackage{bm}
\usepackage{amsmath}
\usepackage{amssymb}
\usepackage[colorlinks=true, allcolors=blue]{hyperref}

\begin{document}
\title{Handling the asymmetric spectral line profile}

\author{D. Solovyev$^{1,2}$} 
\author{A. Anikin$^{1,3}$}
\email[E-mail:]{alexey.anikin.spbu@gmail.com}
\author{T. Zalialiutdinov$^{1,2}$}
\affiliation{ 
$^1$ Department of Physics, St. Petersburg State University, Petrodvorets, Oulianovskaya 1, 198504, St. Petersburg, Russia
\\
$^2$ Petersburg Nuclear Physics Institute named by B.P. Konstantinov of National Research Centre 'Kurchatov Institut', St. Petersburg, Gatchina, 188300, Russia
\\
$^3$D.I. Mendeleev Institute for Metrology, St. Petersburg, 190005, Russia
}

\begin{abstract}
This paper discusses some features of the spectral line profile theory used in the treatment of measured atomic transitions. It is shown that going beyond the established linear approximation for the spectral line contour in the case of its nonresonant extension, the potential for a more accurate extraction of atomic characteristics from experimental data arises. Using the example of the Lyman-$\alpha$ (Ly$_\alpha$) transition in hydrogen, a simple analysis of the observed spectral line distorted by a possible interfering transitions is given. In particular, the results obtained in the present work clearly demonstrate that the processing of the same experimental data at different settings can provide an accurate determination of the transition frequency, the centre of gravity as well as the hyperfine splitting of the ground state in hydrogen-like atomic systems. The latter is especially important for setting up precision spectroscopic experiments on the antihydrogen atom.
\end{abstract}

\maketitle
\section{Introduction}
A detailed theoretical description of the observed spectral line profile has become an inherent part of the precision determination of the transition frequency in the last decade. The first rigorous derivation of the line profile in the framework of quantum electrodynamics (QED) was given in \cite{Low}. It was shown that the line profile is determined by the Lorentz contour in the resonance approximation. Until recently, such approach was sufficient to accurately extract the transition frequency from experimental data, although theoretical studies of the phenomena affecting the line profile have been carried out for quite some time, having emerged as a separate field of research, see, e.g., \cite{Andr,ZSLP-report} and references therein.

In particular, theoretical works dealing with particularities of the line shape have shown the importance of taking into account the asymmetry arising when measuring transition frequencies \cite{LSPS,Jent-Mohr}. Theoretical examination of the spectral line asymmetry unveils a notable distinction in determining the transition frequency for an atom unaffected by the measurement process compared to one influenced by experimental observation. Namely, the frequency obtained in an experiment should be correlated to the Bohr model of the atom (underlying theoretical studies) in order to extract, for example, physical constants or to test fundamental interactions and symmetries.

 The most significant influence comes from the effect resulting from interfering pathways for neighboring states, which is known as the quantum interference effect (QIE) \cite{Jent-Mohr}. Depending on the experimental conditions and method of measuring the transition frequency, the asymmetry of the spectral line profile should be taken into account for each specific experiment \cite{S-2020-importance}. Giving rise to a frequency shift that can significantly exceed, for example, radiative corrections, the corresponding asymmetry is, as usual, calculated theoretically \cite{HH-2010,HH-2011,Brown-2013,Yost_2014,MHH-2015,Amaro-2015,Amaro-mH-2015}. 

The most significant advance was made recently in the \cite{H-exp}. First of all, the experiment \cite{H-exp} demonstrated that the spectral line profile has become a precisely measurable quantity. As another important consequence, it is necessary to point out the processing of the experimental data, which enabled to significantly improve the accuracy of the transition frequency extraction. It is the line profile that represents the physical quantity, while the transition frequency, level widths, etc. are determined as its parameters. The accuracy of determining these characteristics strongly depends on the used fitting contour.

The experimental data treatment performed in \cite{H-exp} was based on the procedure described in \cite{Jent-Mohr} and the use of the Fano contour (Fano-Voigt) to fit the observed line profile, see also \cite{Udem_2019}. Once an indispensable part of the experiment \cite{GM-1s-3s,Mat-2s-8s}, the fit of the observed line beyond the resonance approximation is performed by an asymmetric profile derived in the linear approximation, see \cite{Jent-Mohr} for details. In the lowest order this corresponds exactly to the description of the QIE. With the same result, the QIE can be obtained as the frequency shift via an extremum condition for the line shape given by the corresponding cross section \cite{HH-2010,HH-2011,Brown-2013,Yost_2014,MHH-2015,Amaro-2015,Amaro-mH-2015}. 

Adhering to the analysis presented in \cite{Jent-Mohr,H-exp}, we discuss the principles of extracting the transition frequency from the experiment \cite{Eikema,Pahl-Eikema} and the possibility of accurately determining the hyperfine splitting intervals in the hydrogen atom for the Lyman-$\alpha$ (Ly$_\alpha$) transition. In the framework of the linear approximation, it is shown that a thorough treatment of the experimental data should significantly improve the accuracy of extracting the values of hyperfine intervals in such experiments. The latter may have a key role in the detailed comparison with the antihydrogen atom \cite{Walz_2003,Amole2012,Ahmadi2018}.

\section{Theory of spectral line asymmetry}
\label{theory}

In the Furry picture and $S$-matrix formalism \cite{Furry,Akhiezer,LabKlim} for the one-electron atom the standard Lorentz line profile can be obtained by considering the elastic photon scattering by an atomic electron in the state $n$. In the resonant approximation, the amplitude of the process can be reduced to the form \cite{Low,Andr}:
\begin{eqnarray}
\label{1}
U_{nn}^{\rm sc} = \frac{\langle n |  \boldsymbol{\alpha}\boldsymbol{A}^{*}_{\boldsymbol{k}_1, \boldsymbol{e}_1} | r\rangle \langle r| \boldsymbol{\alpha}\boldsymbol{A}_{\boldsymbol{k}_2, \boldsymbol{e}_2} |n\rangle}{E_r+L_r^{\rm SE}-E_n-\omega-\frac{\mathrm{i}}{2}\Gamma_r},
\end{eqnarray}
where $L_r^{\rm SE}$ and $\mathrm{i}\Gamma_r/2$ represent, respectively, the real and imaginary parts of the one-loop self-energy correction (the leading-order Lamb shift and the level width, respectively), the inserts of which were used to regularize the divergent resonance contribution, see for more details \cite{Low,Andr}. Notations $\boldsymbol{A}_{\boldsymbol{k}_2, \boldsymbol{e}_2}$ and $\boldsymbol{A}^{*}_{\boldsymbol{k}_1, \boldsymbol{e}_1}$ are introduced for the wave functions in the coordinate space representation of absorbed and emitted photons, with the corresponding wave, $\boldsymbol{k}$, and polarization, $\boldsymbol{e}$, vectors, $\boldsymbol{\alpha}$ represents the Dirac matrix, $E_r$ denotes the energy of the resonant state $r$ and $\omega\equiv |\boldsymbol{k}|$ is the frequency of the emitted/absorbed photon (the latter can be redefined by the energy conservation law). The Dirac binding energies are $E_n$ for an arbitrary state of atom $n$, which we characterize by a set of quantum numbers $n l j F$ ($n$ is the principal quantum number, $l$ is the orbital momentum, $j$ represents the total angular momentum of the bound electron, and $F$ represents the total atomic momentum taking into account the nuclear spin). Unless otherwise stated, relativistic units $\hbar=c=m=1$ are used ($\hbar$ is the reduced Planck constant, $c$ is the speed of light, and $m$ is the electron mass).

Then the squared modules of the scattering amplitude (\ref{1}) leads to the standard expression for the Lorentz profile. In the following calculations, we will turn to the dominant effect represented by quantum interference. The QI effect can be obtained by considering neighboring states $r_1$ and $r_2$. Then, the scattering amplitude is
\begin{eqnarray}
\label{2}
U_{nn}^{\rm sc} = \frac{\langle n |  \boldsymbol{\alpha}\boldsymbol{A}^{*}_{\boldsymbol{k}_1, \boldsymbol{e}_1} | r_1\rangle \langle r_1| \boldsymbol{\alpha}\boldsymbol{A}_{\boldsymbol{k}_2, \boldsymbol{e}_2} |n\rangle}{\omega_0-\omega-\frac{\mathrm{i}}{2}\Gamma_{r_1}}+
\\
\nonumber
\frac{\langle n |  \boldsymbol{\alpha}\boldsymbol{A}^{*}_{\boldsymbol{k}_1, \boldsymbol{e}_1} | r_2\rangle \langle r_2| \boldsymbol{\alpha}\boldsymbol{A}_{\boldsymbol{k}_2, \boldsymbol{e}_2} |n\rangle}{\omega_0+\Delta-\omega-\frac{\mathrm{i}}{2}\Gamma_{r_2}},
\end{eqnarray}
where with high accuracy one can set $\Gamma_{r_1}=\Gamma_{r_2}\equiv\Gamma_r$ (as for the Ly$_\alpha$ transition), $\omega_0\equiv E_r-E_n$ (all possible radiative corrections can be included in the definition of $E_n$) and $\Delta = E_{r_2}-E_{r_1}$. The expression (\ref{3}) is written for the case when both resonant lines are observed. It can be considered as excitation by two independent laser beams or a single broadband laser overlapping the two lines.

By introducing a shortcut notation for matrix elements $\langle a |  \boldsymbol{\alpha}\boldsymbol{A}_{\boldsymbol{k}_2, \boldsymbol{e}_2} | r_{1(2)}\rangle \langle r_{1(2)}| \boldsymbol{\alpha}\boldsymbol{A}^{*}_{\boldsymbol{k}_1, \boldsymbol{e}_1} |a\rangle = A_{1(2)}$, for the squared modulus of Eq. (\ref{2}) we arrive at
\begin{eqnarray}
\label{3}
\phi(x) \sim \frac{\left|A_1\right|^2}{x^2+\frac{1}{4}\Gamma^2_r} + \frac{\left|A_2\right|^2}{(x+\Delta)^2+\frac{1}{4}\Gamma^2_r}\qquad
\\
\nonumber
+ \frac{2 Re\left[A_1A^*_2\right]\left[x(x+\Delta)+\frac{1}{4}\Gamma^2_r\right]}{\left[x^2+\frac{1}{4}\Gamma^2_r\right]\left[(x+\Delta)^2+\frac{1}{4}\Gamma^2_r\right]},\qquad
\end{eqnarray}
where $x\equiv \omega_0-\omega$ and $Re[\dots]$ denotes the real part of magnitude in brackets. The dependence on the angles between the polarization and propagation vectors (or their combinations) for the absorbed and emitted photons is contained in the amplitudes $A_1$, $A_2$. The factor $Re\left[A_1A^*_2\right]$ determines the angular dependence of the line profile along with the squares of the corresponding amplitudes.

By applying the extremum condition to Eq.~(\ref{3}), where $d\phi(\omega)/dx=0$, it is possible to derive a fifth-degree equation with respect to $x$. For obvious reasons, the solution of this equation should be performed numerically. However, assuming the smallness of $x$, the conventional manner corresponds to the linear approximation, which immediately leads to an expression for the frequency shift caused by the QIE. The details of such a description are extensively covered in the literature, see, e.g., \cite{AZSL-2022} and references therein. 

Here we only note that the three real roots of the emerging equation can be interpreted, obviously, as values of two maxima and one minimum of the line profile. The latter is of particular interest for determining the "gravity centre" of two lines. According to the results of calculations taking into account the fine structure of the Lyman-$\alpha$ transition, i.e., for the interfering $1s_{1/2}\rightarrow 2p_{1/2}, 2p_{3/2}$ transitions \cite{Jent-Mohr}, this centre of gravity can be shifted by a nonresonant correction. 

To demonstrate this statement, it is enough to consider the case when the interference is zero. Then one can find the magic angle for electric dipole transitions with vanishing amplitude $Re[A_1A^*_2]$ in Eq.~(\ref{3}). By performing the calculations (see Appendix), the roots determined by the extremum condition of Eq.~(\ref{3}) at the magic angle can be found as follows
\begin{eqnarray}
\label{4}
\nonumber
x^{(1)}&=&0, x^{(2)}=\Delta_{\rm fs},
\\
x^{(3)}&=&\left(1-2^{1/3}+2^{2/3}\right)\frac{\Delta_{\rm fs}}{3} -\frac{\left(2-2^{1/3}\right)}{2^{2/3}} \frac{ \Gamma^2}{6 \Delta_{\rm fs}},
\\
\nonumber
x^{(4,5)}&=&0.278753 \Delta_{\rm fs} + 0.0388517\frac{\Gamma^2}{\Delta_{\rm fs}}
\\
\nonumber
&\pm &\mathrm{i} \left(0.296415 \frac{\Gamma^2}{\Delta_{\rm fs}}+0.821951 \Delta_{\rm fs}\right).
\end{eqnarray}
Here $\Delta_{\rm fs}$ denotes the fine splitting interval $E_{2p_{3/2}}-E_{2p_{1/2}}$, and $\Gamma$ is the width of the $2p$ state in the hydrogen atom. The expressions above were obtained for the case of the correlation between the two polarization vectors. The angle between the polarization vector of the absorbed photon and the direction vector of the emitted one will be treated in later sections.

As can be seen from Eqs.~(\ref{4}), the first and second roots correspond to the transition frequencies $E_{1s_{1/2}}-E_{2p_{1/2}}$ and $E_{1s_ {1/2}}-E_{2p_{3/2}}$, respectively. Both are shifted by a fine structure interval with respect to each other, and both are not subject to QIE at a magic angle. The root $x^{(3)}$ represents the position of the minimum of the line profile Eq.~(\ref{3}) and can be understood as the centre of gravity of the two lines, whose position is determined by the coefficient at $\Delta_{\rm fs}$ and is subject to the effect of quantum interference even at a magic angle. 

\begin{figure}[h!]
\centering
 \includegraphics[width=0.9\columnwidth]{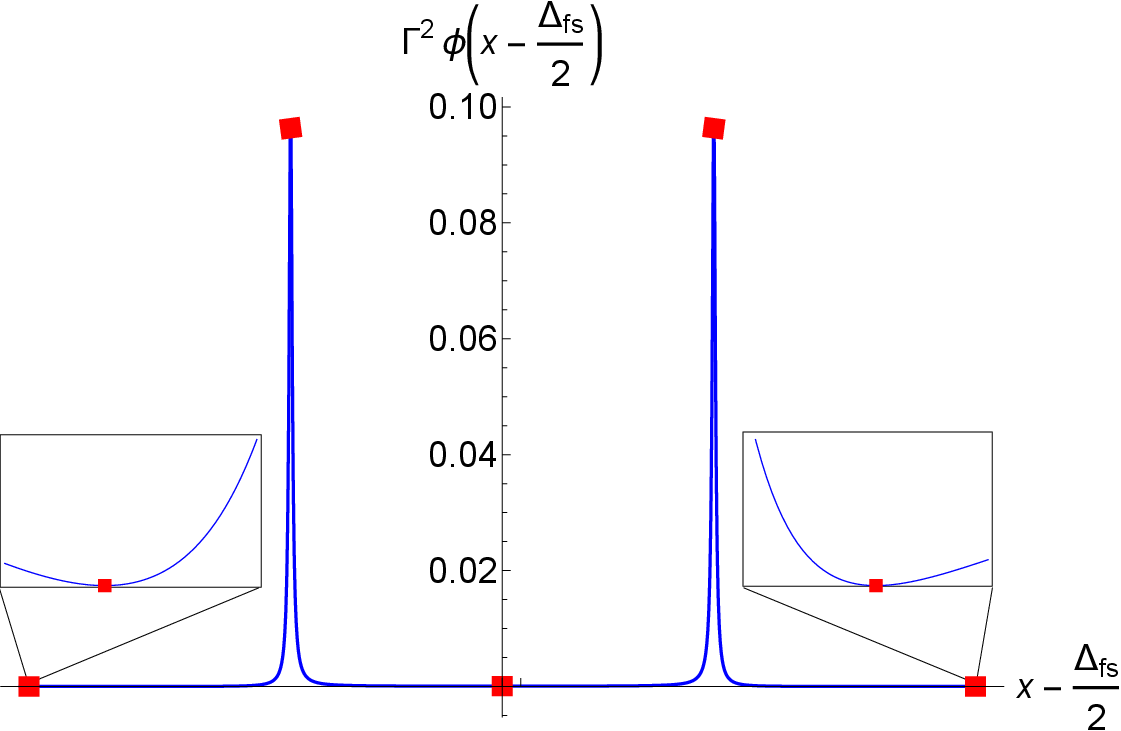}
\caption{Graph illustrating the spectral profile of two lines, $1s_{1/2}\rightarrow 2p_{1/2}$ and $1s_{1/2}\rightarrow 2p_{3/2}$, separated by the fine structure interval $\Delta_{\rm fs}$. Roots corresponding to the extremum condition are marked with red squares. The graph is plotted at an angle of $\pi/2$ between the polarization vectors of the emitted and absorbed photons, and an enlarged region of minima to the left and right of both lines is inserted for demonstration (black arrows connect the same roots). Spectral profile values are multiplied by the square of the natural width for clarity and for convenience, the graph is shifted so that the x-coordinate of the centre of gravity ($\Delta_{\rm fs}/2$) is zero.}
\label{Fig_1}
\end{figure}

Finally, the remaining roots $x^{(4,5)}$ yield inflection points (the second derivative of Eq.~(\ref{3}) is zero at these points). We give their numerical coefficients because of the unwieldiness of the expressions, although we can state that, as in all other roots, the coefficients are determined by the amplitudes of $A_1$ and $A_2$. Keeping in mind that $x=\omega_0-\omega$, the real part of these roots can be interpreted as a shift, and their imaginary part as the value of the width at inflection points (like the natural width in Eqs.~(\ref{1}), (\ref{2})). At first glance, the roots of $x^{(4,5)}$ are unphysical, but depending on the angle, they can become real. Calculations for the angle $\pi/2$ at lowest order give: $x^{(1)}=\Gamma^2/(2\Delta_{\rm fs})$, $x^{(2)} = \Delta_{\rm fs}-\Gamma^2/(2\Delta_{\rm fs})$, $x^{(3)} = \Delta_{\rm fs}/2$ and $x^{(4,5)}=\Delta_{\rm fs}/2\pm\sqrt{5}\Delta_{\rm fs}/2$. The first three roots, as before, represent the maxima (shifted by nonresonant correction) and the unshifted centre of gravity. The centre of gravity is located exactly in the middle of two lines corresponding to fine sublevels at angle $\pi/2$. In turn, the expressions for $x^{(4,5)}$ give additional positions of minima located on the left and right of the two peaks, see Fig.~\ref{Fig_1}.

Thus, measuring the centre of gravity requires different data processing with respect to determining the transition frequency. In the case of a larger number of lines, which may overlap significantly, the appropriate analysis becomes much more complicated (the extremum condition will lead to polynomials of higher powers on $x$).

\section{Determination of the HFS of the ground state of hydrogen by measuring the Ly$_\alpha$ transition}
\label{HFS}

\begin{figure}[h!]
\centering
\includegraphics[width=0.9\columnwidth]{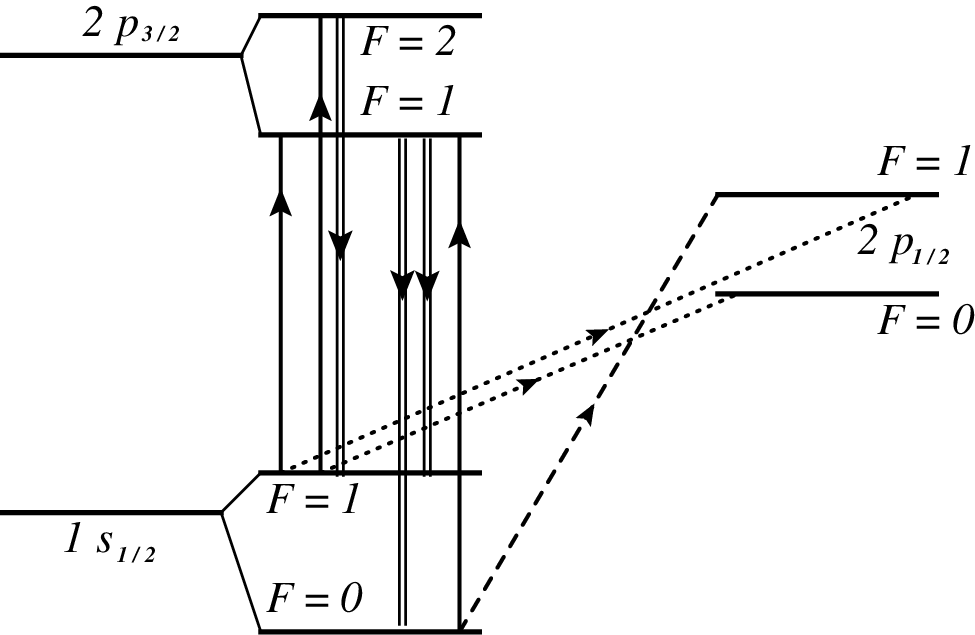}
\caption{Scheme of transitions in the experiment \cite{Eikema}. The hyperfine splittings of levels $1s_{1/2}$, $2p_{3/2}$, as well as the $2p_{1/2}$ state splitting, are denoted by the corresponding values of the total atomic momentum $F$. The transitions are shown by lines with an arrow (single line for absorption and double line for emission). Since there is no direct transition to the $2p_{1/2}$ state, but interference exists, the corresponding pathways are shown by dashed and dotted lines.
}
\label{Fig_2}
\end{figure}

We now turn to the linear approximation, within which the line profile fitting was presented theoretically in \cite{Jent-Mohr} and later used in \cite{H-exp} to treat the experimental results. Considering the Lyman-$\alpha$ transition, we discuss the possibility of accurately determining the hyperfine structure (HFS) interval of the ground state in the hydrogen atom. Referring to the experiment \cite{Eikema}, the level scheme is shown in Fig.~\ref{Fig_2}.

The experiment \cite{Eikema} showed that the observation of Lyman-$\alpha$ emission lines can be used to accurately measure the hyperfine structure (HFS) of the ground state. A value differing by $\approx 20$ MHz from the generally accepted result (see, for example, \cite{Hellwig1970,Diermaier2017}) was found in order to further apply a similar scheme to measurements in antimatter \cite{Eikema}. Possessing particular importance, a detailed comparison of the hyperfine ground state of hydrogen and antihydrogen atoms was recently performed in \cite{Ahmadi2017}, where the HFS interval for the latter was determined as $1420.4\pm 0.5$ MHz. It should be noted that hydrogen HFS interval experiments are orders of magnitude more accurate than antihydrogen experiments, see for example \cite{Hellwig1970,Diermaier2017}. Although experiments with the antihydrogen atom are more challenging (see also \cite{Ahmadi2018}), we turn to the experimental setup of the \cite{Eikema}, suggesting their future adaptation to antimatter. The main purpose of the following discussion is to demonstrate that the HFS interval determination of the ground state through Ly$_\alpha$ emission can be determined more accurately using the same experiment as \cite{Eikema}.

To realize this, one should consider the emission lines (depicted by the double solid line in Fig.~\ref{Fig_2}) without distinguishing the hyperfine sublevels of the initial state, but for different hyperfine sublevels of the final state, i.e., with total atomic momenta $F_f=1$ and $F_f=0$. In consistency with the experiment \cite{Eikema}, one should also sum over the total momentum, $F$, of the $2p_{3/2}$ resonance state atom, since excitation is possible to any hyperfine sublevel. Then, the scattering amplitude can be presented in form:
\begin{eqnarray}
\label{5}
U_{1s_{1/2}\rightarrow 2p_{3/2}\rightarrow 1s_{1/2}^{F=0}}+U_{1s_{1/2}\rightarrow 2p_{3/2}\rightarrow 1s_{1/2}^{F=1}} = 
\nonumber
\\
\frac{A_1}{E_{2p_{3/2}}-\omega-E_{1s_{1/2}^{F=1}}-\mathrm{i}\frac{\Gamma}{2}}+\frac{A_2}{E_{2p_{1/2}}-\omega-E_{1s_{1/2}^{F=1}}}+
\\
\nonumber
\frac{A_3}{E_{2p_{3/2}}-\omega-E_{1s_{1/2}^{F=0}}-\mathrm{i}\frac{\Gamma}{2}}+\frac{A_4}{E_{2p_{1/2}}-\omega-E_{1s_{1/2}^{F=0}}}.
\end{eqnarray}
Here the divergent resonance contributions have been regularized according to the procedure presented in \cite{Low}. The terms with amplitudes $A_2$ and $A_4$ were introduced as interfering paths to the $2p_{1/2}$ fine splitting state. The notations for amplitudes in Eq.~\ref{5} are
\begin{eqnarray}
\label{6}
A_1=\sum\limits_{F_i}(2F_i+1)\sum\limits_{F}\langle 1s_{1/2}^{F_i}|\boldsymbol{\alpha}\boldsymbol{A}^{*}_{\boldsymbol{k}_1, \boldsymbol{e}_1} | 2p_{3/2}^F \rangle\times
\\
\nonumber
\langle 2p_{3/2}^F | \boldsymbol{\alpha}\boldsymbol{A}_{\boldsymbol{k}_2, \boldsymbol{e}_2}|1s_{1/2}^{F_f=1}\rangle,
\\
\nonumber
A_2=\sum\limits_{F_i}(2F_i+1)\sum\limits_{F}\langle 1s_{1/2}^{F_i}|\boldsymbol{\alpha}\boldsymbol{A}^{*}_{\boldsymbol{k}_1, \boldsymbol{e}_1} | 2p_{1/2}^F \rangle\times
\\
\nonumber
\langle 2p_{1/2}^F | \boldsymbol{\alpha}\boldsymbol{A}_{\boldsymbol{k}_2, \boldsymbol{e}_2}|1s_{1/2}^{F_f=1}\rangle,
\\
\nonumber
A_3=\sum\limits_{F_i}(2F_i+1)\sum\limits_{F}\langle 1s_{1/2}^{F_i}|\boldsymbol{\alpha}\boldsymbol{A}^{*}_{\boldsymbol{k}_1, \boldsymbol{e}_1} | 2p_{3/2}^F \rangle\times
\\
\nonumber
\langle 2p_{3/2}^F | \boldsymbol{\alpha}\boldsymbol{A}_{\boldsymbol{k}_2, \boldsymbol{e}_2}|1s_{1/2}^{F_f=0}\rangle,
\\
\nonumber
A_4=\sum\limits_{F_i}(2F_i+1)\sum\limits_{F}\langle 1s_{1/2}^{F_i}|\boldsymbol{\alpha}\boldsymbol{A}^{*}_{\boldsymbol{k}_1, \boldsymbol{e}_1} | 2p_{1/2}^F \rangle\times
\\
\nonumber
\langle 2p_{1/2}^F | \boldsymbol{\alpha}\boldsymbol{A}_{\boldsymbol{k}_2, \boldsymbol{e}_2}|1s_{1/2}^{F_f=0}\rangle.
\end{eqnarray}
The $(2F_i+1)$ coefficient is introduced so that the excitation process proceeds in accordance with the experiment \cite{Eikema}, from the $1s_{1/2}$ state with an unfixed hyperfine sublevel.

By performing the calculations, see \cite{S-2020-importance} for details, we find in the nonrelativistic limit
\begin{eqnarray}
\label{7}
\nonumber
\left|A_1\right|^2=-\frac{1}{54}\left(7-3\cos{2\vartheta}\right),
\\
\left|A_2\right|^2=-\frac{2}{27},
\\
\nonumber
A_1 A_2=\frac{1}{54}\left(1+3\cos{2\vartheta}\right)
\end{eqnarray}
for the decay to the $1s_{1/2}^{F_f=1}$ state, and for the $1s_{1/2}^{F_f=0}$ state:
\begin{eqnarray}
\label{8}
\nonumber
\left|A_3\right|^2=-\frac{1}{162}\left(7-3\cos{2\vartheta}\right),
\\
\left|A_4\right|^2=-\frac{2}{81},
\\
\nonumber
A_3 A_4=\frac{1}{162}\left(1+3\cos{2\vartheta}\right).
\end{eqnarray}
Here we assume that there is no interference between channels with different total atomic momentum of the final state, i.e. between transitions with $F_f=1$ and $F_f=0$.

Presence of interference in Eqs.~(\ref{7}) and (\ref{8}) results in asymmetric line profiles for two scattering channels $1s_{1/2}\rightarrow 2p_{3/2}\rightarrow 1s_{1/2}^{F_f=1}$ and $1s_{1/2}\rightarrow 2p_{3/2}\rightarrow 1s_{1/2}^{F_f=0}$. The corresponding transition frequencies can be determined accurately in two ways. The first is given by subtracting the nonresonant correction from the transition frequency, determined as the maximum of the line profile. This correction in the linear approximation on $x$ is easily calculated and is expressed through $\Gamma^2/(4\Delta_{\rm fs})\approx -227.4$ kHz, where $\Delta_{\rm fs}$ is the fine structure interval between states $2p_{3/2}$ and $2p_{1/2}$. At the same time, as before, it follows from Eqs.~(\ref{7}) and (\ref{8}) that the angle at which the low-order asymmetry vanishes is the magic angle, $\vartheta_{\rm m}=\arccos{(1/\sqrt{3})}=\arccos{(-1/3)}/2$. This is the result of distinguishable transitions into hyperfine sublevels of the ground state. 

Another scenario refers to the use of an asymmetric profiles to fit the observed spectral line shapes. Following the theory stated in \cite{Jent-Mohr}, within a linear approximation the fitting contour can be given by a sum of two profiles:
\begin{eqnarray}
\label{9}
\frac{C_1}{(x-\Delta_1(x))^2+\frac{1}{4}\Gamma^2} + \frac{C_2}{(x+\Delta^{(1s)}_{\rm HFS}-\Delta_2(x))^2+\frac{1}{4}\Gamma^2}.\qquad
\end{eqnarray}
In Eq.~(\ref{9}) the first function is written for the scattering channel to the $1s_{1/2}^{F_f=1}$ state and the second function corresponds to $1s_{1/2}^{F_f=0}$. The introduced coefficients are 
\begin{eqnarray}
\label{10}
C_1=\left|A_1\right|^2,\, a_1 = 2\frac{\left|A_2\right|^2}{\Delta_{\rm fs}^3},\, b_1=-2\frac{A_1 A_2}{\Delta_{\rm fs}}
\\
\nonumber
C_2=\left|A_3\right|^2,\, a_2= 2\frac{\left|A_4\right|^2}{\Delta_{\rm fs}^3},\, b_2=-2\frac{A_3 A_4}{\Delta_{\rm fs}}.
\end{eqnarray}
The frequency shifts are defined as
\begin{eqnarray}
\label{11}
\Delta_1(x) = \frac{a_1}{2C_1}\left[x^2+\frac{1}{4}\Gamma^2\right]^2 + \frac{b_1}{2C_1}\left[x^2+\frac{1}{4}\Gamma^2\right],
\\
\nonumber
\Delta_2(x) = \frac{a_2}{2C_2}\left[\left(x+\Delta^{(1s)}_{\rm HFS}\right)^2+\frac{1}{4}\Gamma^2\right]^2
\\
\nonumber
 + \frac{b_2}{2C_2}\left[\left(x+\Delta^{(1s)}_{\rm HFS}\right)^2+\frac{1}{4}\Gamma^2\right].
\end{eqnarray}

\begin{figure}[hbtp]
\centering
\includegraphics[width=0.9\columnwidth]{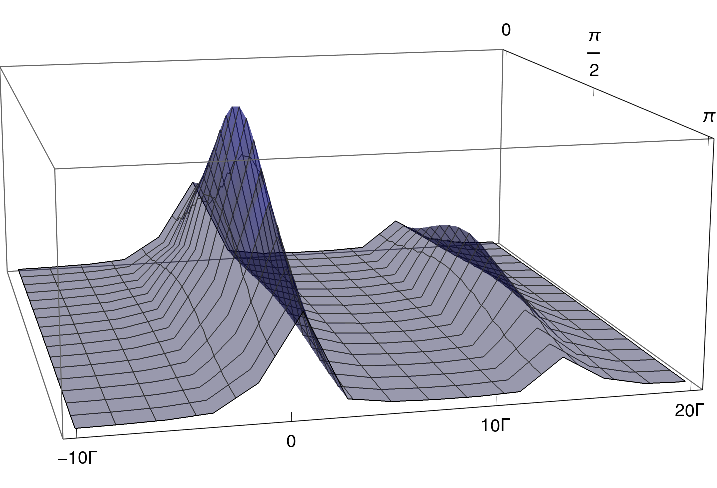}
\caption{The spectral profile arising for two scattering channels $1s_{1/2}\rightarrow 2p_{3/2}\rightarrow 1s_{1/2}^{F=1}$ and $1s_{1/2}\rightarrow 2p_{3/2}\rightarrow 1s_{1/2}^{F=0}$ as a function of frequency and angle in arbitrary units for visualisation. The axis with notations in gamma refers to the variable $x\equiv E_{2p_{3/2}^{F=1}}-E_{1s_{1/2}^{F=1}}-\omega$, the axis in $\pi$ scale refers to the dependence of line profile on $\vartheta$ angle, and the line contour values are plotted along the height. 
Two resonant lines occur at $x=0$ and $x=\Delta^{(1s)}_{\rm HFS}$.}
\label{Fig_3}
\end{figure}

Thus, the contour given by the expression (\ref{9}) becomes defined as a function of the angle $\vartheta$ and the atomic characteristics $\Delta_{\rm fs}$, $\Delta^{(1s)}_{\rm HFS}$, $\Gamma$. The corresponding profile is shown in Fig.~\ref{Fig_3} as a function of frequency and angle. The difference between the model shape (\ref{9}) and the one based on (\ref{5}) is invisible to the naked eye.

Solving equation $\Delta_1(x=0)=0$ (here $x=\omega_0-\omega \equiv E_{2p_{3/2}}-E_{1s_{1/2}^{F_f=1}}-\omega$) and $\Delta_2\left(x=-\Delta^{(1s)}_{\rm HFS}\right)=0$, one can find the angle at which the unshifted frequencies can be determined. The result is
\begin{eqnarray}
\label{12}
\vartheta_0=\pm\frac{1}{2}\arccos{\left[\frac{-2\Gamma^2-\Delta_{\rm fs}^2}{3\Delta_{\rm fs}^2}\right]},
\end{eqnarray}
which gives the "magic angle" in approximation of small width $\Gamma$. The addition to the magic angle at lowest order is $\pm \Gamma^2/\left(2\sqrt{2}\Delta_{\rm fs}^2\right)\approx 2.9\times 10^{-5}$ in radians. 

So far we have considered the case of determining the hyperfine splitting interval by finding the two frequencies corresponding to the $1s_{1/2}\rightarrow 2p_{3/2}\rightarrow 1s_{1/2}^{F=1 }$ and $1s_{1/2}\rightarrow 2p_{3/2}\rightarrow 1s_{1/2}^{F=0}$ spectral lines. Accordingly, there are two possibilities: i) to determine the frequencies and angle from experimental data using the asymmetric profile (\ref{9}), ii) to utilize the symmetric line profile obtained by fitting parameter equal to the angle $\vartheta=\vartheta_0$ calculated theoretically. Such a symmetric contour will not neatly fit the observed line shape, but the maxima will be unshifted, which in turn can be used to define $\Delta^{(1s)}_{\rm HFS}$ as the frequency difference at their maxima (symmetrization procedure \cite{H-exp}). 

However, according to the previous section, the hyperfine splitting of the ground state can be found in another way. The centre of gravity can be used for this purpose. Solving the set of equations $\Delta_1(x)=0$, $\Delta_2(x)=0$, one can find ten pairs $\{x,\vartheta\}$, two of which are real. Dropping out the solutions corresponding to $x=\pm\mathrm{i}\Gamma/2$ and  $x=-\Delta^{(1s)}_{\rm HFS}\pm\mathrm{i}\Gamma/2$, we found
\begin{eqnarray}
\label{13}
x=-\frac{\Delta^{(1s)}_{\rm HFS}}{2}
\end{eqnarray}
at the angle
\begin{eqnarray}
\label{14}
\vartheta_{\rm gc} = \pm \frac{1}{2}\arccos{\left(\frac{-\Delta^2_{\rm fs} - 2\left(\Delta^{(1s)}_{\rm HFS}\right)^2 - 2\Gamma^2}{3\Delta_{\rm fs}^2}\right)}.
\end{eqnarray}
The angle $\vartheta_{\rm gc}\approx 0.961319$ and differs on the magic angle, $\vartheta_0$: $\delta\vartheta =\vartheta_{\rm gc}-\vartheta_0 \approx \left(\Delta^{(1s)}_{\rm HFS}\right)^2/\left(2\sqrt{2}\Delta_{\rm fs}^2\right)\approx 6\times 10^{-3}$. However, at angle $\vartheta_{\rm gc}$ the centre of gravity is unshifted and is located exactly (in lowest order) in the middle of the two peaks, thus providing an accurate determination of the hyperfine splitting of the ground state.

\section{$2p_{3/2}$ HFS in Ly$_\alpha$ transition measurements}
\label{2pHFS}

So far we have described the Ly$_\alpha$ transition without emphasizing the hyperfine structure of the $2p_{3/2}$ state, $\Delta_{\rm HFS}^{(2p_{3/2})}$. Meanwhile, the value of the appropriate energy interval is about $23.6516$ MHz, see \cite{HH-tab}, which admits the influence of the $2p_{3/2}$ state hyperfine structure on the determination of the HFS of the ground state carried out in \cite{Eikema}. Note that the value of the $\Delta_{\rm HFS}^{(2p_{3/2})}$ splitting is about four times smaller than the natural level width $\Gamma\approx 99.7624$ MHz, which leads to the failure to observe it in experiments of the type. Hence, investigating the impact of such a blurring on the determination of the $\Delta^{(1s)}_{\rm HFS}$, as well as the possibility of extracting the $\Delta_{\rm HFS}^{(2p_{3/2})}$ interval, is the main goal of this section.

To account for the $2p_{3/2}$ level hyperfine splitting of atomic hydrogen, we use the experimental setup of \cite{Eikema}, see Fig. ~\ref{Fig_2}, where two emission lines corresponding to $F_f =0$ and $F_f=1$ of the $1s_{1/2}$ ground state were observed. Then, evidently, the line corresponding to the final momentum $F_f=0$ is not responsible for the $\Delta_{\rm HFS}^{(2p_{3/2})}$, and moreover does not interfere with the channel scattering to $F_f=0$. So, we start by considering the scattering channel $1s_{1/2}\rightarrow 2p_{3/2}^{F=1}/2p_{3/2}^{F=2}\rightarrow 1s_{1 /2}^{F_f=1}$, which can be further generalized to the case described in the previous section~\ref{HFS} (by adding the remaining line profile in Eq.~(\ref{9})). 

According to the expression (\ref{2}), we have
\begin{eqnarray}
\label{15}
U^{\rm sc}_{1s_{1/2}\rightarrow 1s_{1/2}^{F_f=1}} \sim \frac{A_1}{x-\frac{\mathrm{i}}{2}\Gamma}+\frac{A_2}{x+\Delta_{\rm HFS}^{(2p_{3/2})}-\frac{\mathrm{i}}{2}\Gamma}\qquad
\\
\nonumber
+\frac{A_3}{x-\Delta_{\rm fs}^{(11)}}+\frac{A_4}{x-\Delta_{\rm fs}^{(10)}}+\frac{A_5}{x-\Delta_{\rm fs}^{(21)}}+\frac{A_6}{x-\Delta_{\rm fs}^{(20)}}.
\end{eqnarray}
Here the first two terms represent transitions to $2p_{3/2}^{F=1}$ and $2p_{3/2}^{F=2}$, respectively, while the last four terms correspond to the interferening paths with $2p_{1/2}^{F=1}$ and $2p_{1/2}^{F=0}$ separated by the fine splitting interval. The corresponding splitting energies are $\Delta_{\rm fs}^{(11)}=E_{2p_{3/2}^{F=1}}-E_{2p_{1/2}^{F=1}}$, $\Delta_{\rm fs}^{(10)} = E_{2p_{3/2}^{F=1}}-E_{2p_{1/2}^{F=0}}$, $\Delta_{\rm fs}^{(21)}=E_{2p_{3/2}^{F=2}}-E_{2p_{1/2}^{F=1}}$, $\Delta_{\rm fs}^{(20)} = E_{2p_{3/2}^{F=2}}-E_{2p_{1/2}^{F=0}}$, and $x\equiv E_{2p_{3/2}^{F=1}}-E_{1s_{1/2}^{F_f=1}}-\omega$.

At first, it can be found that the resonant terms give rise to the overlapping profiles (indistinguishable due to the level width). The interference represented by the product $A_1A_2$ cannot be attributed to either one of them. However, using the approximation of $\Delta_{\rm HFS}^{(2p_{3/2})}$ smallness, we can write
\begin{eqnarray}
\label{16}
\phi^{(0)}(x)\approx \frac{\left|A_1\right|^2+A_1A_2}{x^2+\frac{1}{4}\Gamma^2}+\frac{\left|A_2\right|^2}{\left(x+\Delta_{\rm HFS}^{(2p_{3/2})}\right)^2+\frac{1}{4}\Gamma^2},\qquad
\end{eqnarray}
where the contribution $\sim -\Delta_{\rm HFS}^{(2p_{3/2})}\left(x+\Delta_{\rm HFS}^{(2p_{3/2})}\right)$ is discarded. Performing calculations as before, we obtain
\begin{eqnarray}
\label{17}
\left|A_1\right|^2 = -\frac{1}{648}(17+ 3\cos{2 \vartheta}),
\nonumber
\\
\left|A_2\right|^2 = \frac{1}{648}(21 \cos{2 \vartheta}-73),
\\
\nonumber
A_1A_2 = \frac{1}{216} (1 + 3 \cos{2\vartheta}).
\end{eqnarray}
The result (\ref{17}) shows that albeit $\left|A_1\right|^2$ and $\left|A_2\right|^2$ cannot be zero, the first term vanishes when $\left| A_1 \right|^2+2 A_1A_2=0$, i.e. $\left|A_1\right|^2+2A_1A_2 = (15\cos(2\vartheta)-11)/648=0$ and hence $\vartheta=\pm (1/2)\arccos(11/15)\approx 21.4^\circ$.

\begin{figure}[hbtp]
\centering
\includegraphics[width=0.9\columnwidth]{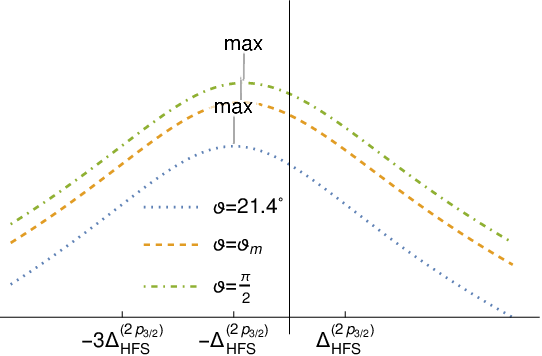}
\caption{The spectral line profile given by Eq.~(\ref{16}). To illustrate the shift of the profile maximum, the graph is plotted in logarithmic scale for the values of angle $\vartheta = (1/2)\arccos(11/15)$, magic angle $\vartheta=\arccos(1/ \sqrt{3})$, and $\vartheta=\pi/2$.}
\label{Fig_4}
\end{figure}

The shift of the maximum as a function of angle is shown in Fig.~\ref{Fig_4}.
In spite of obvious shift of the line profile maximum from Fig.~\ref{Fig_4}, this representation does not take into account the asymmetry caused by interference transitions due to the fine structure of levels. The latter can be considered with the remaining terms in Eq.~(\ref{15}) independently. 

According to the theory of \cite{Jent-Mohr}, generalized in the previous section to the two observed lines, one should evaluate the amplitudes and determine the asymmetry parameters. Then the spectral profile of the emission line corresponding to the final state $1s_{1/2}^{F_f=1}$ is given by
\begin{eqnarray}
\label{18}
\phi(x)\approx 
\frac{\left|A_1\right|^2+A_1A_2}{(x-\tilde{\Delta}_1(x))^2+\frac{1}{4}\Gamma^2}
\\
\nonumber
+\frac{\left|A_2\right|^2}{\left(x+\Delta_{\rm HFS}^{(2p_{3/2})} - \tilde{\Delta}_2(x)\right)^2+\frac{1}{4}\Gamma^2}.
\end{eqnarray}
Here $\tilde{\Delta}_1(x)$ and $\tilde{\Delta}_2(x)$ are expressed as before, see Eq. (\ref{11}), with the substitution $\Delta^{(1s)}_{\rm HFS}\rightarrow \Delta_{\rm HFS}^{(2p_{3/2})}$ and coefficients
\begin{eqnarray}
\label{19}
\tilde{C}_1=\left|A_1\right|^2+A_1A_2,\, 
\tilde{a}_1 &=& \frac{2\left|A_3\right|^2}{\left(\Delta_{\rm fs}^{(11)}\right)^3}+\frac{2\left|A_4\right|^2}{\left(\Delta_{\rm fs}^{(10)}\right)^3},
\nonumber
\\
\tilde{b}_1 &=& -\frac{2A_1A_3}{\Delta_{\rm fs}^{(11)}}-\frac{2A_1A_4}{\Delta_{\rm fs}^{(10)}},
\\
\nonumber
\tilde{C}_2=\left|A_2\right|^2,\, 
\tilde{a}_2 &=& \frac{2\left|A_5\right|^2}{\left(\Delta_{\rm fs}^{(21)}\right)^3}+\frac{2\left|A_6\right|^2}{\left(\Delta_{\rm fs}^{(20)}\right)^3},
\nonumber
\\
\nonumber
\tilde{b}_2 &=& -\frac{2A_2A_5}{\Delta_{\rm fs}^{(21)}}-\frac{2A_2A_6}{\Delta_{\rm fs}^{(20)}},
\end{eqnarray}
and together with the expressions (\ref{17}) we found
\begin{eqnarray}
\label{20}
\left|A_3\right|^2 = -\frac{4}{81},\, A_1A_3 = \frac{1+3\cos{2\vartheta}}{324},
\nonumber
\\
\left|A_4\right|^2 = -\frac{2}{81},\, A_1A_4 = 0,
\\
\nonumber
\left|A_5\right|^2 = -\frac{4}{81},\, A_2A_5 = \frac{1+3\cos{2\vartheta}}{108},
\\
\nonumber
\left|A_6\right|^2 = -\frac{2}{81},\, A_2A_6 = \frac{1+3\cos{2\vartheta}}{162}.
\end{eqnarray}

Taking into account the result (\ref{17}), we find that the emission line profile to the $1s_{1/2}^{F_f=1}$ state can be approximated by the second summand in Eq. (\ref{18}), where $\left|A_2\right|^2$ is defined at angle $(1/2)\arccos(11/15)$. As a next step to accurately consider the hyperfine splitting of the $2p_{3/2}$ state, an angle close to the magic angle should obviously be chosen. Numerical calculation gives $\tilde{\Delta}_2(x=0)=0$ at $\vartheta=\pm 0.955338$ ($\vartheta_{\rm m}=0.955317$).

\begin{figure}[h!]
\centering
\includegraphics[width=0.9\columnwidth]{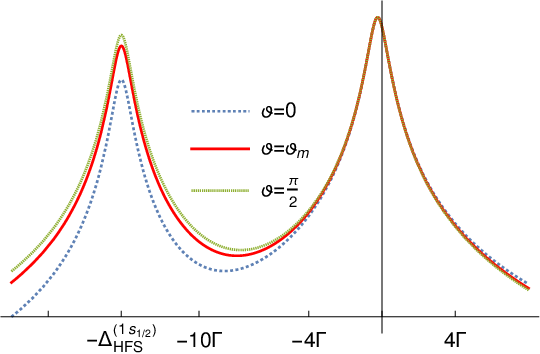}
\caption{The line profile constructed as the second summands in Eqs. (\ref{9}), (\ref{18}). The graphs are plotted in logarithmic scale at different angles $\vartheta=0$, $\vartheta=\vartheta_{\rm m}$ and $\vartheta=\pi/2$ to clearly demonstrate the difference in decay emission line to the $1s_{1/2}^{F=0}$ state, while the second emission line remains the same.}
\label{Fig_5}
\end{figure}

\begin{figure}[h!]
\centering
\includegraphics[width=0.9\columnwidth]{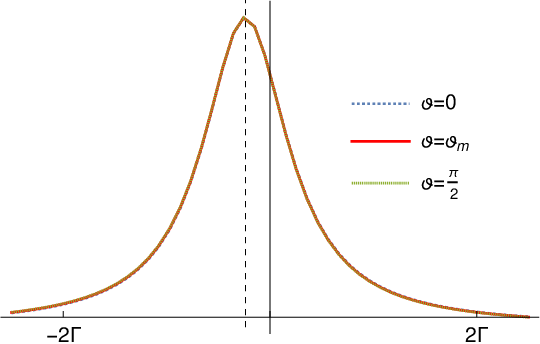}
\caption{The emission line to the $1s_{1/2}^{F=1}$ state at different angles $\vartheta=0$, $\vartheta=\vartheta_{\rm m}$ and $\vartheta=\pi/2$. At arbitrary angles, the profiles coincide due to the smallness of the parameters $\tilde{a}_2$ and $\tilde{b}_2$. Dashed line is plotted at $x=-\Delta_{\rm HFS}^{(2p_{3/2})}$.}
\label{Fig_6}
\end{figure}

The schematic behavior of the profile constructed as the second summands in Eqs. (\ref{9}) and (\ref{18}) is illustrated in Fig.~\ref{Fig_5}.
In particular, from Fig.~\ref{Fig_5} it is clear that with the chosen asymmetry parameters, the emission line to the $1s_{1/2}^{F=1}$ state remains unchanged. In turn, the line profile corresponding to the state $1s_{1/2}^{F=0}$ changes significantly. Moreover, a shift in the local minimum (centre of gravity) is also obvious. Finally, in Fig.~\ref{Fig_6} we show in detail the emission line to the $1s_{1/2}^{F=1}$ state.
The corresponding line profile is shifted by the value $\Delta_{\rm HFS}^{(2p_{3/2})}$.

\section{Conclusions}
\label{concl}

In this work, we analyzed the solutions of the equation arising from the extremum condition determined for the spectral line profile. Here, for simplicity, the extremum condition has been used to determine the transition frequency of the observed spectral line (the full-width-half-maximum condition, i.e., $x=\Gamma/2$, can just as well be used). In the case of an 'isolated' line, there is a single solution corresponding to the line profile parameter representing the transition frequency. However, the presence of a neighboring lines leads to significant complications. The asymmetry of the line profile arising from nonresonant contributions to the scattering cross section should be taken into account.

The most considerable contribution to the line asymmetry comes from the interference effect occurring for the two pathways between neighboring resonance states. As a rule for the description of QIE, the linear approximation for the line profile is employed (Fano contour). The QI effect has been widely discussed in the literature and has several important properties. One of them is that the angular dependence is determined by the set of quantum numbers for the described transitions. 

Another important aspect of the QIE is that the dominant nonresonant contribution can be reduced to zero by choosing the angle between the absorbed and emitted photons. For the hydrogen atom, a theoretical description of the one-photon scattering process was given in \cite{LSPS,Jent-Mohr} and later adapted in \cite{H-exp} to accurately determine the $2s-4p$ transition frequency. In particular, it turns out that the magic angle $\vartheta_{\rm m}=\arccos(1/\sqrt{3})$ should be used for electric dipole transitions. At the same time, the model profile of the line (Fano contour, see \cite{Jent-Mohr}) taking into account the asymmetry parameter demonstrates an improved extrapolation of the experimental data to beyond the resonance approximation, as in \cite{H-exp}.

Using the Ly$_\alpha$ transition as an example, we find that the choice of the magic angle is very relevant for determining the unshifted transition frequency and fine splitting interval (and hence the second transition frequency). However, the centre of gravity of these lines cannot be defined at this angle through an integer relation, such as a centroid for example. Determination of the gravity centre is subjected to nonresonant effects in another way, the angle dependence is determined not by the ratio of resonant and nonresonant amplitudes but by their permissible combinations. The physical interpretation of all roots for the equation arising from the extremum condition clearly demonstrates this. The two lines corresponding to fine sublevels of the Lyman-$\alpha$ transition (discarding the hyperfine structure of levels) are equally distant from the centre of gravity at an angle of $\pi/2$, although each is distorted due to nonresonant effects.

Despite the standalone interest in the gravity centre of lines in atomic spectroscopy, its measurement can also be used to accurately determine the hyperfine splitting. According to the analysis of section~\ref{HFS}, we at first considered the possibility of processing experimental data to determine the transition frequencies to selected hyperfine sublevels of the ground state in the hydrogen atom, discarding the hyperfine structure of the $2p_{3/2}$ state. Then in section~\ref{2pHFS} the HFS of the resonant state was included into analysis. The main conclusion of this consideration is that the hyperfine splitting interval can be found exactly as the difference of the transition frequencies representing the maxima of these symmetrized lines. 

For this purpose, the symmetrization procedure should be performed using Eqs.~(\ref{9}), (\ref{18}) in accordance with the theory \cite{Jent-Mohr}. It can be expected that the HFS interval of the ground state can be determined more accurately than it was found in \cite{Eikema} from exactly the same experiment. In principle, determining the HFS of the $2p_{3/2}$ state is also possible (although it is four times smaller than the natural level width) by noticing the shift of the maximum at different angles and calculating the frequency shift in accordance with the extremum condition. Focusing on $\Delta^{(1s)}_{\rm HFS}$, according to the conclusions drawn in sections~\ref{theory}, \ref{HFS} and \ref{2pHFS}, we find that first the emission line corresponding to the transition to $1s_{1/2}^{F=1}$ state should be processed to obtain a symmetrized profile shifted by the value of $\Delta_{\rm HFS}^{(2p_{3/2})}$. The angle determining the corresponding asymmetry parameters is about $21.4^\circ$. Then other asymmetry parameters (determined at the magic angle) should be used to avoid QIE due to the fine structure of the levels.

The importance of the above analysis can be emphasized by recent experiments \cite{Ahmadi2018} aimed at a detailed comparison of the spectra of hydrogen and antihydrogen. It can be expected that the results obtained in this work can serve for the experimental improvement of such measurements.

\section*{Acknowledgements}
The work of T. Z. and A. A. was supported by the foundation for the advancement of theoretical physics and mathematics ”BASIS” (grants No. 23-1-3-31-1 and 22-1-5-9-1, respectively).

\bibliography{mybibfile}

\appendix
\renewcommand{\theequation}{A\arabic{equation}}
\setcounter{equation}{0}
\setcounter{figure}{0}
\renewcommand{\thefigure}{A\arabic{figure}}
\renewcommand{\thetable}{A\arabic{table}}
\renewcommand{\bibnumfmt}[1]{[S#1]}
\renewcommand{\citenumfont}[1]{S#1}

\onecolumngrid
\section{Evaluation of transition matrix elements}
\label{Ap1}
\begin{figure}[hbtp]
\centering
\includegraphics[width=0.7\columnwidth]{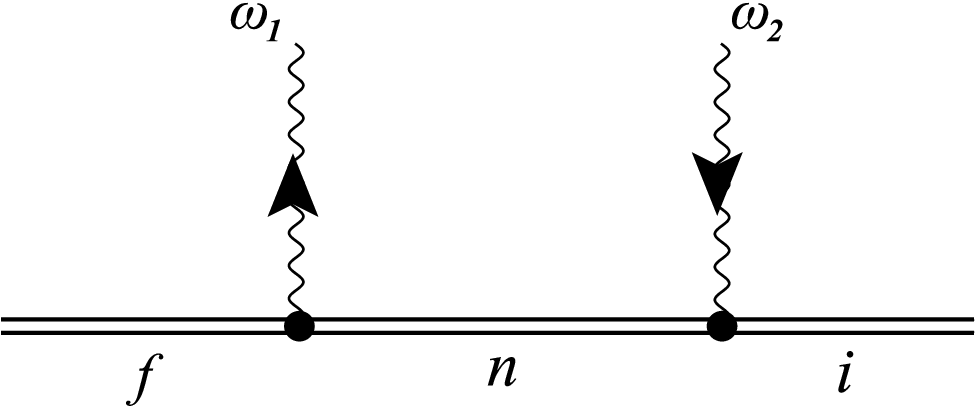}
\caption{Feynman diagram representing the one-photon scattering process by an atomic electron. Wavy lines with arrows denote absorption (down arrow) and emission (up arrow) with corresponding frequencies $\omega_2$ and $\omega_1$, respectively. The double solid line represents the bound electron state (Furry picture). The designations $f$, $n$ and $i$ denote the final, intermediate, and initial states.}
\label{figS1}
\end{figure}

To describe the one-photon scattering process, we use the $S$-matrix formalism \citep{Akhiezer,Greiner}. According to the Feynman diagram depicted in Fig.~\ref{figS1},
 the $S$-matrix element is
\begin{eqnarray}
\label{S.1}
S_{fi}=(-\mathrm{i}e)^2\int d^4x_1 d^4x_2 \overline{\psi}_f(x_1)\gamma^\mu A_\mu^*(x_1)
S(x_1, x_2)\gamma^\nu A_\nu(x_2)\psi_i(x_2).
\end{eqnarray} 
The graph in Fig.~\ref{figS1} should additionally include a part with permuted photons. However, this part is not essential for our purposes, and we omit its consideration (the corresponding discussion can be found, for example, in \cite{LKG,LSPS}).

For an arbitrary atomic state $a$
\begin{eqnarray}
\label{S.2}
\psi_a(x)=\psi_a(\boldsymbol{r})e^{-\mathrm{i}E_at},
\end{eqnarray}
where $\psi_a(\boldsymbol{r})$ is the solution of the Dirac equation for the atomic electron, $E_a$ is the Dirac energy, $\overline{\psi}_a=\psi^+_a\gamma_0$ is the Dirac conjugated wave function, $\gamma_{\mu}\equiv(\gamma_0,\boldsymbol{\gamma})$ are the Dirac matrices and $x\equiv(t,\boldsymbol{r})$ (vectors are in bold) is the four-dimensional space-time coordinate. 
The photon field or the photon wave function $A_{\mu}(x)$ is defined by
\begin{eqnarray}
\label{S.3}
A_{\mu}(x)=\sqrt{\frac{2\pi}{\omega}}e_{\mu}e^{\mathrm{i}(\boldsymbol{k}\boldsymbol{r}-\omega t)}=e^{-\mathrm{i}\omega t}A_{\mu}(\boldsymbol{r}),
\end{eqnarray}
where $e_{\mu}$ are the components of the photon polarization four-vector ($\boldsymbol{e}$ is 3-dimensional polarization vector for real photons), $k\equiv(\omega,\boldsymbol{k})$ is the photon momentum four-vector, $\boldsymbol{k}$ is the wave vector, $\omega=|\boldsymbol{k}|$ is the photon frequency. Eq. (\ref{S.3}) corresponds to the absorbed photon and $A^*_{\mu}(x)$ corresponds to the emitted photon. Finally, the electron propagator for the bound electron can be presented in the form of the eigenmode decomposition with respect to one-electron eigenstates:
\begin{eqnarray}
\label{S.4}
S(x_1,x_2)=\frac{\mathrm{i}}{2\pi}\int\limits^{\infty}_{-\infty}d\Omega\, e^{-i\Omega(t_1-t_2)}\sum\limits_{n}\frac{\psi_n(\boldsymbol{r}_1)\overline{\psi}_n(\boldsymbol{r}_2)}{\Omega-E_n(1-\mathrm{i}0)}.
\end{eqnarray}

Integration over time variables yields
\begin{eqnarray}
\label{S.5}
S_{fi} = -2\pi\mathrm{i}e^2\int d\boldsymbol{r}_1 d\boldsymbol{r}_2 \psi_f^*(\boldsymbol{r}_1)\boldsymbol{\alpha}\boldsymbol{A}^{*}_{\boldsymbol{k}_1, \boldsymbol{e}_1}\int\limits^{\infty}_{-\infty}d\Omega\, \sum\limits_{n}\frac{\psi_n(\boldsymbol{r}_1)\overline{\psi}_n(\boldsymbol{r}_2)}{\Omega-E_n(1-\mathrm{i}0)}\boldsymbol{\alpha}\boldsymbol{A}_{\boldsymbol{k}_1, \boldsymbol{e}_1}\psi_i(\boldsymbol{r}_2)\delta(E_f+\omega_1-\Omega)\delta(\Omega-\omega_2-E_i),\qquad
\end{eqnarray}
where $\boldsymbol{\alpha}$ is the Dirac matrix and $\boldsymbol{A}_{\boldsymbol{k}, \boldsymbol{e}}$ represents the vector part of the photon wave function characterized by the wave vector $\boldsymbol{k}$ and transversal polarization $\boldsymbol{e}$. Then, after the integration over $\Omega$, we have
\begin{eqnarray}
\label{S.6}
S_{fi} = -2\pi\mathrm{i}e^2\delta(E_f+\omega_1-\omega_2-E_i)\sum\limits_n\frac{\langle f| \boldsymbol{\alpha}\boldsymbol{A}^{*}_{\boldsymbol{k}_1, \boldsymbol{e}_1} | n\rangle\langle n| \boldsymbol{\alpha}\boldsymbol{A}_{\boldsymbol{k}_1, \boldsymbol{e}_1} | i\rangle}{E_i+\omega_2-E_n(1-\mathrm{i}0)}.
\end{eqnarray}
For the equal initial and final states (elastic scattering), $f=i=a$ and $\omega_1=\omega_2\equiv\omega$, using the relation $S_{fi}=-2\pi \mathrm{i}\,U_{fi}$, the expression (\ref{S.6}) gives the amplitude of the scattering process:
\begin{eqnarray}
\label{S.7}
U_{aa}^{\rm sc} = e^2\sum\limits_n\frac{\langle a| \boldsymbol{\alpha}\boldsymbol{A}^{*}_{\boldsymbol{k}_1, \boldsymbol{e}_1} | n\rangle\langle n| \boldsymbol{\alpha}\boldsymbol{A}_{\boldsymbol{k}_1, \boldsymbol{e}_1} | a\rangle}{E_a+\omega-E_n(1-\mathrm{i}0)},
\end{eqnarray}
which in the resonance approximation (the resonant state is fixed in the sum over $n$), $n=r$, reduces to
\begin{eqnarray}
\label{S.8}
U_{aa}^{\rm sc} =e^2\frac{\langle a| \boldsymbol{\alpha}\boldsymbol{A}^{*}_{\boldsymbol{k}_1, \boldsymbol{e}_1} | r\rangle\langle r| \boldsymbol{\alpha}\boldsymbol{A}_{\boldsymbol{k}_1, \boldsymbol{e}_1} | a\rangle}{E_a+\omega-E_r(1-\mathrm{i}0)}.
\end{eqnarray}

\begin{figure}[hbtp]
\centering
\includegraphics[scale=0.7]{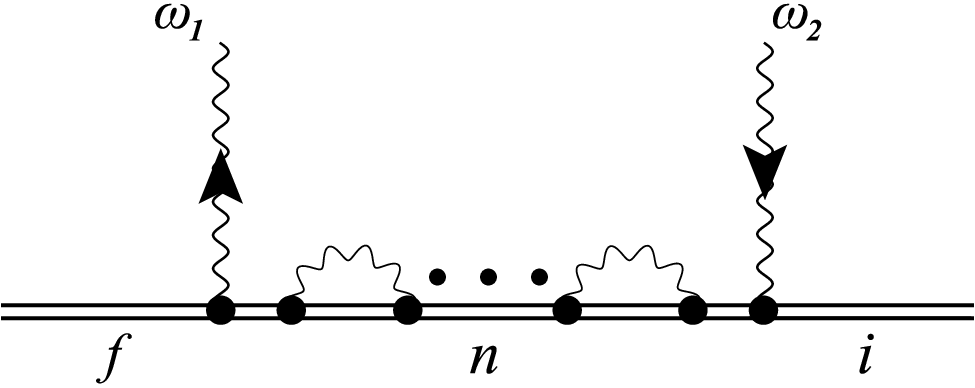}
\caption{Feynman diagram representing the one-loop self-energy insertions into the photon scattering amplitude. All designations are the same as in Fig.~\ref{figS1}. The parts given by the photon loop (closed wavy lines) are the corresponding lower order self-energy correction. The dots denote an infinite number of such inserts. The summation of the diagrams with a different number of loops is assumed.}
\label{figS2}
\end{figure}

To obtain the spectral line profile, it is necessary to regularize the expression (\ref{S.8}), since it diverges at $\omega=E_r-E_a$. This can be done using the procedure described in \cite{Low}, see also \cite{Andr} for details, according to which an infinite number of one-loop self-energy corrections should be inserted sequentially into the graph in Fig.~\ref{figS1}. The regularization procedure is shown in Fig~\ref{figS2}.

The sum of the diagrams in Fig.~\ref{figS2} is a geometric progression, which leads to the appearance of the real and imaginary parts of the lower-order self-energy correction in the energy denominator of Eq. (\ref{S.8}), see \cite{Andr}:
\begin{eqnarray}
\label{S.9}
U_{aa}^{\rm sc} = e^2\frac{\langle a| \boldsymbol{\alpha}\boldsymbol{A}^{*}_{\boldsymbol{k}_1, \boldsymbol{e}_1} | r\rangle\langle r| \boldsymbol{\alpha}\boldsymbol{A}_{\boldsymbol{k}_1, \boldsymbol{e}_1} | a\rangle}{E_a+\omega-E_r-L_r^{\rm SE}+\mathrm{i}\frac{\Gamma_r}{2}},
\end{eqnarray}
where we have used that the imaginary part of the self-energy correction is minus half of the $\Gamma_r$ level width, and the real part is the lower-order Lamb shift of the $r$ resonance state.
 In turn, choosing the nearest states in the sum over $n$, one can write
\begin{eqnarray}
\label{S.10}
U_{aa}^{\rm sc} = e^2\frac{\langle a| \boldsymbol{\alpha}\boldsymbol{A}^{*}_{\boldsymbol{k}_1, \boldsymbol{e}_1} | r_1\rangle\langle r_1| \boldsymbol{\alpha}\boldsymbol{A}_{\boldsymbol{k}_1, \boldsymbol{e}_1} | a\rangle}{E_a+\omega-E_{r_1}-L_{r_1}^{\rm SE}+\mathrm{i}\frac{\Gamma_{r_1}}{2}} + 
e^2\frac{\langle a| \boldsymbol{\alpha}\boldsymbol{A}^{*}_{\boldsymbol{k}_1, \boldsymbol{e}_1} | r_2\rangle\langle r_2| \boldsymbol{\alpha}\boldsymbol{A}_{\boldsymbol{k}_1, \boldsymbol{e}_1} | a\rangle}{E_a+\omega-E_{r_2}-L_{r_2}^{\rm SE}+\mathrm{i}\frac{\Gamma_{r_2}}{2}},
\end{eqnarray} 
where we used the regularization procedure depicted in Fig.~\ref{figS2}. Introducing the notation $\omega_0=E_{r_1}-E_a$, we find that the second energy denominator is shifted by $\Delta= E_{r_2}-E_{r_1}$ with respect to the first one. 

The cross section of the elastic one-photon scattering process is obtained by the formula:
\begin{eqnarray}
\label{S.11}
d\sigma^{\rm sc}_a = 2\pi\sum\limits_{\boldsymbol{e}_1\boldsymbol{e}_1}\left|U_{aa}^{sc}\right|^2\delta(\omega_1-\omega_2)\frac{d\boldsymbol{k}}{(2\pi)^3},
\end{eqnarray}
which should be summed over the projection of the angular momentum of the final state and averaged over the projection of the initial state. The summation over the projections of the intermediate states in Eqs. (\ref{S.9}) and (\ref{S.10}) is also assumed. The $\delta$-function is written for clarity in Eq. (\ref{S.11}) and was taken into account in the expressions (\ref{S.7})-(\ref{S.10}).

Further evaluation concerns the matrix elements in the numerators of Eq. (\ref{S.10}). For the hydrogen atom within the framework of the dipole approximation $e^{i\boldsymbol{k}\boldsymbol{r}}\approx 1$ and $\boldsymbol{\alpha}\boldsymbol{A}_{\boldsymbol{k}, \boldsymbol{e}} = \sqrt{2\pi/\omega}\,\boldsymbol{e}\boldsymbol{\alpha}$. Then, in the nonrelativistic limit $\boldsymbol{\alpha}$ is replaced by $\boldsymbol{p}$ (electron momentum operator). The electron momentum operator can be replaced by the coordinate, $\boldsymbol{r}$, according to the commutation relation $\boldsymbol{p}=\mathrm{i}[\hat{H},\boldsymbol{r}]$, which for the matrix element $(\boldsymbol{p})_{ab}$ reduces to $\mathrm{i}(E_a-E_b)(\boldsymbol{r})_{ab}=\Delta E_{ab}(\boldsymbol{r})_{ab}$. According to this, we arrive at
\begin{eqnarray}
\label{S.12}
U_{aa}^{\rm sc} = e^2 2\pi \omega_0^2\left[\frac{\langle a| \boldsymbol{e}_2\boldsymbol{r} | r_1\rangle\langle r_1| \boldsymbol{e}^*_1\boldsymbol{r} | a\rangle}{\omega_0-\omega - \mathrm{i}\frac{\Gamma_{r_1}}{2}} + 
e^2\frac{\langle a| \boldsymbol{e}^*_1\boldsymbol{r} | r_2\rangle\langle r_2| \boldsymbol{e}_2\boldsymbol{r} | a\rangle}{\omega_0-\omega +\Delta - \mathrm{i}\frac{\Gamma_{r_2}}{2}}\right].
\end{eqnarray}
Here the wave functions are given by the solution of the Schr\"{o}dinger-Pauli equation. The common pre-factor arising before the squared amplitude is not essential for our purposes and, in further, we focus on the matrix elements in brackets of Eq. (\ref{S.12}). 

Considering the states shifted relative to each other by the fine structure, the resonance states $r_1$ and $r_2$ are described as the atomic energy levels with different angular momentum of the bound electron, $j$. Then, the states can be characterized by the set of quantum numbers $njlsm_j$, where $n$ is the principal quantum number, $l$ is the orbital momentum, $s$ represents the spin of the bound electron and $j$, $m_j$ represents the total angular momentum, $\boldsymbol{j}=\boldsymbol{l}+\boldsymbol{s}$ and its projection, respectively. Using the definition of the scalar product $(\boldsymbol{e}\boldsymbol{r})=\sum_q(-1)^q\boldsymbol{e}_q\boldsymbol{r}_{-q}$, the matrix elements in Eq. (\ref{S.12}) can be evaluated via the Wigner-Eckart theorem:
\begin{eqnarray}
\label{S.13}
\langle n'j'm'_j|\hat{T}_{k\kappa}|njm_j\rangle = (-1)^{2k}C_{jm_j\,k\kappa}^{j'm'_j}\frac{\langle n'j'\|\hat{T}_k\|nj\rangle}{\sqrt{2j'+1}},
\end{eqnarray}
where the coefficient $C_{j_1m_1\,j_2m_2}^{jm}$ gives the Clebsch-Gordan coefficient for the decomposition of $|jm\rangle$ in terms of $|j_1m_1\rangle$, $|j_2m_2\rangle$ \cite{Varsh}. The reduced matrix element of the spin-independent operator for the set of quantum numbers $njlsm_j$ is defined by
\begin{eqnarray}
\label{S.14}
\langle n'l's'j'\|\hat{T}_k\| nlsj\rangle = \delta_{ss'}(-1)^{j+l'+s+k}\Pi_{jj'}
\begin{Bmatrix}
l & s & j \\
j' & k & l'
\end{Bmatrix}
\langle n'l'\|\hat{T}_k\|nl\rangle.
\end{eqnarray}
Then, for the coordinate operator
\begin{eqnarray}
\label{S.15}
\langle n'l'\|r\|nl\rangle = (-1)^{l'}\Pi_{ll'}
\begin{pmatrix}
l' & 1 & l \\
0 & 0 & 0
\end{pmatrix}
\int\limits_0^\infty dr\, r^3\, R_{n'l'}(r)R_{nl}(r),
\end{eqnarray}
where $R_{nl}(r)$ is the radial part of the hydrogen wave function.

The summation over projections can be performed as follows. For the squared amplitude (\ref{S.12}) we should consider
\begin{eqnarray}
\label{S.16}
\left|A\right|^2 = \sum\limits_{\substack{q_1 q_2\\ q'_1 q'_2}}(-1)^{q_1+q_2+q'_1+q'_2}\boldsymbol{e}^*_{1_{q_1}}\boldsymbol{e}_{2_{q_2}}\boldsymbol{e}_{1_{q'_1}}\boldsymbol{e}^*_{2_{q'_2}}
   \sum\limits_{\substack{m_{j_f}m_{j_i}\\m_{j_r}m'_{j_r}}}\langle n_a l_a s j_a m_{j_f}|r_{-q_1}|n_r l_r s j_r m_{j_r}\rangle\langle n_r l_r s j_r m_{j_r}|r_{-q_2}|n_a l_a s j_a m_{j_i}\rangle    \qquad
\nonumber
\\
  \times\langle n_a l_a s j_a m_{j_i}|r_{-q'_2}|n_r l_r s j'_r m'_{j_r}\rangle\langle n_r l_r s j'_r m'_{j_r}|r_{-q'_1}|n_a l_a s j_a m_{j_i}\rangle,\qquad\qquad
\end{eqnarray}
where it is taken into account that for the process of elastic scattering, the initial and final states can differ in the projection of the angular momentum and for a one-electron atom $s_a=s_r\equiv s=1/2$ with fixed orbital momenta of the initial, final, and resonant states. Denoting all coefficients which are independent of projections as $\{ {\rm red}\}$, we find that the angular correlations between the polarization vectors are determined by
\begin{eqnarray}
\label{S.17}
\left|A\right|^2 = \sum\limits_{\substack{q_1 q_2\\ q'_1 q'_2}}(-1)^{q_1+q_2+q'_1+q'_2}\boldsymbol{e}^*_{1_{q_1}}\boldsymbol{e}_{2_{q_2}}\boldsymbol{e}_{1_{q'_1}}\boldsymbol{e}^*_{2_{q'_2}}
   \sum\limits_{\substack{m_{j_f}m_{j_i}\\m_{j_r}m'_{j_r}}} \{ {\rm red} \} \Pi^{-1}_{j_a j_r j_a j'_r} C^{j_am_{j_f}}_{j_rm_r\,1-q_1}\, C^{j_rm_{j_r}}_{j_am_{j_i}\,1-q_2}\, C^{j_am_{j_i}}_{j'_rm'_r\,1-q'_2}\, C^{j'_rm'_{j_r}}_{j_am_{j_f}\, 1-q'_1}.\qquad
\end{eqnarray}

By rearranging the indices of the Clebsch-Gordan coefficients according to the symmetry property \cite{Varsh}, we have employed the following relation:
\begin{eqnarray}
\label{S.18}
\sum\limits_{\delta}C^{e\epsilon}_{b\beta\, d\delta}C^{d\delta}_{a\alpha\, f\varphi} = \sum\limits_{c\gamma}(-1)^{2e}\Pi_{cd}C^{c\gamma}_{a\alpha\,b\beta}C^{e\epsilon}_{f\varphi\, c\gamma}
\begin{Bmatrix}
a & b & c \\
e & f & d
\end{Bmatrix}.
\end{eqnarray}
Then, summing over $m_{j_r}$ and $m'_{j_r}$, one can obtain
\begin{eqnarray}
\label{S.19}
\left|A\right|^2 \sim  \sum\limits_{\substack{q_1 q_2\\ q'_1 q'_2}}(-1)^{q_1+q_2+q'_1+q'_2}\boldsymbol{e}^*_{1_{q_1}}\boldsymbol{e}_{2_{q_2}}\boldsymbol{e}_{1_{q'_1}}\boldsymbol{e}^*_{2_{q'_2}}
\sum\limits_{m_{j_f}m_{j_i}}    \sum\limits_{c\gamma\, d\delta}C^{c\gamma}_{1-q_2\,1-q_1}C^{j_am_{j_f}}_{j_am_{j_i}\,c\gamma}C^{d\delta}_{1-q'_1\,1-q'_1}C^{j_am_{j_i}}_{j_am_{j_f}\, d\delta}.
\end{eqnarray}
Summing over $m_{j_f}$ and $m'_{j_i}$, we can get
\begin{eqnarray}
\label{S.20}
C^{j_am_{j_f}}_{j_am_{j_i}\,c\gamma}C^{j_am_{j_i}}_{j_am_{j_f}\, d\delta} = (-1)^{-\gamma}\frac{2j_a+1}{2c+1}\delta_{c\,d}\delta_{\gamma\,-\delta}.
\end{eqnarray}
Then, taking into account the symmetry property of the Clebsch-Gordan coefficients and $q_1+q_2+q'_1+q'_2 = -\gamma-\delta$, by the definition of the tensor product
\begin{eqnarray}
\label{S.21}
\left|A\right|^2 \sim \sum\limits_{c\gamma}(-1)^{c-\gamma}\{ \boldsymbol{e}_2\otimes\boldsymbol{e}^*_1 \}_{c-\gamma} \{ \boldsymbol{e}^*_2\otimes\boldsymbol{e}_1 \}_{c\gamma}.
\end{eqnarray}
The latter is the scalar product of two tensors.

Performing the numerical calculations for Eq. (\ref{S.21}), one can find
\begin{eqnarray}
\label{S.22}
\left|A_{j_r=j'_r=\frac{1}{2}}\right|^2 &\equiv& \left|A_1\right|^2 = \frac{2 R}{81}\left(3\{ \boldsymbol{e}_2\otimes\boldsymbol{e}^*_1 \}_0 \{ \boldsymbol{e}^*_2\otimes\boldsymbol{e}_1 \}_0 + 2\{ \boldsymbol{e}_2\otimes\boldsymbol{e}^*_1 \}_1 \{ \boldsymbol{e}^*_2\otimes\boldsymbol{e}_1 \}_1\right)=\frac{2 R}{81}\left(\cos^2\vartheta+\sin^2\vartheta\right) = \frac{2 R}{81},\qquad
\nonumber
\\
A_{j_r=\frac{1}{2}}A_{j'_r=\frac{3}{2}} &\equiv& A_1A_2 = \frac{4R}{81}\left(3\{ \boldsymbol{e}_2\otimes\boldsymbol{e}^*_1 \}_0 \{ \boldsymbol{e}^*_2\otimes\boldsymbol{e}_1 \}_0 - \{ \boldsymbol{e}_2\otimes\boldsymbol{e}^*_1 \}_1 \{ \boldsymbol{e}^*_2\otimes\boldsymbol{e}_1 \}_1\right) = \frac{2 R}{81}\left(2\cos^2\vartheta-\sin^2\vartheta\right),
\\
\nonumber
\left|A_{j_r=j'_r=\frac{3}{2}}\right|^2 &\equiv& \left|A_2\right|^2 =  \frac{4 R}{81}\left(6\{ \boldsymbol{e}_2\otimes\boldsymbol{e}^*_1 \}_0 \{ \boldsymbol{e}^*_2\otimes\boldsymbol{e}_1 \}_0 + \{ \boldsymbol{e}_2\otimes\boldsymbol{e}^*_1 \}_1 \{ \boldsymbol{e}^*_2\otimes\boldsymbol{e}_1 \}_1\right)=\frac{2 R}{81}\left(4\cos^2\vartheta+\sin^2\vartheta\right).
\end{eqnarray}
Here $R$ denotes all the remaining radial integrals, and we have used relations for the tensor components of ranks $0$, $1$ with scalar and vector products of polarization vectors \cite{Varsh}. The second equation can be solved with respect to $\vartheta$: $2\cos^2\vartheta-\sin^2\vartheta = 3\cos^2\vartheta-1=0$ leading to the magic angle $\vartheta_{\rm m} = \arccos(1/\sqrt{3})$ at which the interference of two pathways $j_a=1/2\rightarrow j_r=1/2$ and $j_a=1/2\rightarrow j'_r=3/2$ is equal to zero.


\end{document}